\documentclass{aa}
\usepackage{epsfig}
\usepackage{natbib}
\bibpunct{(}{)}{;}{}{,}

\newif\ifAMStwofonts
\def\spose#1{\hbox to 0pt{#1\hss}}
\def\simlt{\mathrel{\spose{\lower 3pt\hbox{$\mathchar"218$}}
     \raise 2.0pt\hbox{$\mathchar"13C$}}}
\def\simgt{\mathrel{\spose{\lower 3pt\hbox{$\mathchar"218$}}
     \raise 2.0pt\hbox{$\mathchar"13E$}}}

\def\etal{{\rm et~al. }}
\def\hi{H\,{\sc i}~}
\def\oi{O\,{\sc i}~}
\def\ni{N\,{\sc i}~}
\def\ari{Ar\,{\sc i}~}
\def\hii{H\,{\sc ii}~}

\begin{document}
\title{Continuous star formation in IZw18}
\author{Simone Recchi\inst{1, 2}\thanks{recchi@mpa-garching.mpg.de} 
        \and Francesca Matteucci\inst{3}\thanks{matteucci@ts.astro.it} 
        \and Annibale D'Ercole\inst{4}\thanks{annibale@bo.astro.it} 
        \and Monica Tosi\inst{4}\thanks{tosi@bo.astro.it}}
\offprints{S. Recchi}
\institute{
Institut f\"ur theoretische Physik und Astrophysik, Kiel University, 
Olshausenstrasse 40, D-24118 Kiel, Germany \and
Max-Planck Institut f\"ur Astrophysik, 
Karl-Schwarzschild-Strasse 1, D-85741 Garching bei M\"unchen, Germany \and
Dipartimento di Astronomia, Universit\`a di Trieste, Via G.B. Tiepolo, 11,
34131 Trieste, Italy \and
INAF - Osservatorio Astronomico di Bologna, Via Ranzani 1, 40127
Bologna, Italy}
\date{Received  /  Accepted   }

\abstract{ We study the dynamical and chemical evolution of a galaxy
similar to IZw18 under the assumption of a continuous star formation
during bursts.  We adopt a 2-D hydrocode coupled with detailed
chemical yields originating from SNeII, SNeIa and from single
intermediate-mass stars.  Different nucleosynthetic yields and
different IMF slopes are tested.  In most of the explored cases, a
galactic wind develops, mostly carrying out of the galaxy the
metal-enriched gas produced by the burst itself.  The chemical species
with the largest escape probabilities are Fe and N.  Consequently, we
predict that the [$\alpha$/Fe] and [$\alpha$/N] ratios outside the
galaxy are lower than inside.  In order to reproduce the chemical
composition of IZw18, the best choice seems to be the adoption of the
yields of Meynet \& Maeder (2002) which take into account stellar
rotation, although these authors do not follow the whole evolution of
all the stars.  Models with a flat IMF (x=0.5) seem to be able to
better reproduce the chemical properties of IZw18, but they inject in
the gas a much larger amount of energy and the resulting galactic wind
is very strong, at variance with observations.  We also predict the
evolution of the abundances in the \hi medium and compare them with
recent {\sl FUSE} observations.

\keywords{Hydrodynamics -- ISM: abundances -- ISM: jets and outflows --
Galaxies: evolution -- Galaxies: individual: IZw18}}
\maketitle
\bigskip\bigskip

\section{Introduction}

Blue Compact Dwarf (BCD) galaxies, due to their very blue colors,
compact appearance, high gas content and low metallicities, are
generally thought to be unevolved systems.  With an oxygen abundance
(O/H) $\sim$ 0.02 (O/H)$_\odot$, IZw18 (Mrk 116) is the BCD galaxy
with the lowest known metallicity as measured by nebular emission
lines and the ideal candidate for a truly young local galaxy.  This
very low metal content was first observed by Searle and Sargent
(1972), then confirmed by many studies (e.g. Lequeux et al. 1979;
French 1980; Kinman \& Davidson 1981; Kunth \& Sargent 1983; Davidson
\& Kinman 1985; Davidson, Kinman \& Friedman 1989; Pagel et al. 1992;
Skillman \& Kennicutt 1993; Thuan, Izotov \& Lipovetsky 1995; Legrand
et al. 1997; V\'ilchez \& Iglesias-P\'aramo 1998).  Since its
discovery (Zwicky 1966; Sargent \& Searle 1970), this galaxy has been
extensively studied and nowadays we know with reasonable accuracy the
chemical composition of the \hii regions (see e.g. Skillman \&
Kennicutt 1993; Garnett et al. 1995; Izotov \& Thuan 1999 among
others).  IZw18 is thus a good benchmark to study the impact of
starburst(s) on the ISM and compare the results of models with
detailed metallicity determinations.

The determination of the metallicity in the \hi regions is much more
difficult, since the \oi line at 1302~\AA~\ is already strongly
saturated at N$_{H I}$ $\simgt$ 2 $\cdot$ 10$^{20}$ cm$^{-2}$ (the
column density of IZw18 is larger than 10$^{21}$ cm$^{-2}$; van Zee et
al. 1998).  The first determination of \hi metallicity content made by
Kunth et al. (1994) (an abundance in the \hi medium $\sim$ 30 times
lower than the metallicity of the \hii regions) was therefore
uncertain and has been disputed by Pettini \& Lipman (1995).  Recent
{\sl FUSE} (Far Ultraviolet Spectroscopic Explorer) data (Aloisi \etal
2003; Lecavelier des Etangs \etal 2004) allowed observations of \oi
lines at shorter wavelengths and with smaller oscillation strengths.
The knowledge of the metallicity of the neutral medium has therefore
been improved, but uncertainties are still present.  Aloisi \etal
(2003) and Lecavelier des Etangs \etal (2004), analyzing the same {\sl
FUSE} data, obtained different results.  Aloisi \etal (2003) found
abundances of various elements a factor of 3 -- 10 lower than those
found in the \hii regions, with the exception of iron, whose abundance
is the same.  Lecavelier des Etangs \etal (2004) instead found an
oxygen abundance consistent with the one found in the \hii regions,
whereas N and Ar are found to be much more underabundant.

The study of starburst and BCD galaxies by means of hydrodynamical
simulations has been performed by several authors in the recent past.
The overall picture is that galactic winds, produced by the combined
effect of SNe explosions and stellar winds, are not able to eject a
significant fraction of the ambient Interstellar Medium (ISM), since,
due to the flattened distribution of the gas, the wind develops only
in the polar direction (where the pressure gradient is steeper),
whereas the gas in the disc is almost unperturbed.  The final fate of
the metals is still debated.  Many authors (De Young \& Heckman 1994;
D'Ercole \& Brighenti 1999; MacLow \& Ferrara 1999) believe that the
metals are easily channelled along the funnel created by the galactic
wind, so they are ejected outside the galaxy more easily than the
pristine gas.  Other authors (Silich \& Tenorio-Tagle 1998; Legrand et
al. 2001) have suggested that the metal-rich gas is barely lost from
the galaxies, since it is trapped by extended halos surrounding the
galaxies, although the existence of such halos is still unproved.
Rieschick \& Hensler (2000) have suggested that $\sim$ 25 \% of the
metals can mix rapidly, remaining trapped in the galaxy, whereas the
other 75 \% of metals undergoes a cycle lasting $\sim$ 1 Gyr, in which
an outflow is created.  Cooling creates condensation of the metal-rich
gas into molecular droplets, able to fall back and settle onto the
disc of the galaxy.

These studies do not focus on any specific object, trying to study
general properties of starburst and BCD galaxies.  There are a few
exceptions: Strickland \& Stevens (1999) for NGC 5253 and Strickland
\& Stevens (2000) for M82 in particular.  More recently, Vorobyov et
al. (2004) simulated the expansion of supershells in a model galaxy
resembling Holmberg I.  These authors tried to constrain in particular
the gravitational potential and the number of SNeII needed to
reproduce the observed \hi morphology and extension.  Models with a
detailed chemical treatment and with a comparison of the predicted
chemical evolution with the observed metallicity of the target galaxy
are still missing in the literature.

In two previous papers (Recchi, Matteucci \& D'Ercole 2001,
hereafter RMD; Recchi et al. 2002, hereafter R02), we explored the
possibility of a star formation occurring in instantaneous burst(s) in
a model galaxy resembling IZw18. The main results of these two papers
are:

\begin{enumerate}

\item the abundances of heavy elements observed in IZw18 can be
reproduced either with a single burst model, with an age of 31 -- 35
Myr, depending on the model, or with two instantaneous bursts of star
formation separated by a quiescent period of 300 -- 500 Myr.  The
favoured age of the youngest generation of stars is  4 -- 7 Myr,
depending on the models.  A second solution (with an age of several
tens of Myr) is present, but it cannot be accepted since it does not
reproduce the colors and the spectral energy distribution observed in
IZw18. 

\item A galactic wind develops in almost all the models, as a
consequence of the energy supplied by SNe (of both Type Ia and Type
II) and stellar winds.  This wind carries away mostly the metals
freshly produced during the burst.

\item SNeIa explode in a medium already heated and diluted by the
previous activity of SNeII, thus they can hardly loose energy due to
radiative processes.  Owing to that, the galactic wind is mostly
triggered by SNeIa and the metals produced by this kind of SNe (mainly
iron-peak elements) are ejected more easily than the products of SNeII
(mostly $\alpha$-elements).  The [$\alpha$/Fe] ratios outside the
galaxy are therefore lower than inside.

\item The cooling of metals in the gas has been found to be more
efficient than expected and most of the freshly produced metals are
already found in a cold phase (i.e. with temperatures below 2 $\cdot$
10$^4$ K) after 10 -- 12 Myr (but see Sect. 5.1.2 for a critical
discussion about this point).  This can justify the so-called
``instantaneous mixing'' approximation, commonly assumed in chemical
evolution models (Matteucci 1996 and references therein).

\end{enumerate}

In this paper, we propose to relax the hypothesis of instantaneous
bursts of star formation and analyze models with a more complex star
formation regime, as suggested by fitting the observed Color-Magnitude
Diagram (CMD) with synthetic ones (Aloisi, Tosi \& Greggio 1999;
hereafter ATG).  In Sect. 2 we summarize the current literature claims
about the age and the past SFH of IZw18; in Sect. 3 we describe the
model and the assumptions adopted in our simulations.  In Sect. 4 we
present our results.  We discuss them in Sect. 5 and draw some
conclusion in Sect. 6.

\section{The history of the star formation history determinations of IZw18}

IZw18 is a very blue galaxy ($U-B=-0.88$; van Zee et al. 1998),
originally described by Zwicky (1966) as a double system of compact
galaxies.  Following studies have shown that these two components are
two compact superclusters within the same galaxy, separated by 5''.8
(Papaderos et al. 2002).  The two star forming regions are associated
with two \hii regions: the NW and the SE region.  Both regions are
embedded in a low surface brightness envelope, extending to a radius
of $\sim$ 1 Kpc (Dufour \& Hester 1990; Martin 1996) and a much more
extended \hi envelope, which has a size of several Kpc and a total
mass of $\sim$ 6 $\cdot$ 10$^7$ M$_\odot$ (van Zee et al. 1998), but
only $\sim$ 2 $\cdot$ 10$^7$ M$_\odot$ of \hi are associated with the
optical part of the galaxy (Lequeux \& Viallefond 1980; van Zee et
al. 1998).  The total \hii mass is of the order of 3 $\cdot$ 10$^6$
M$_\odot$ (Dufour \& Hester 1990).  The optical part of the galaxy is
often referred to as the {\it Main Body}.  Hereafter we identify this
region as the {\it Galactic Region}.

Historically, IZw18 has been considered as a very young galaxy,
probably experiencing star formation for the first time (Searle \&
Sargent 1972; Hunter \& Thronson 1995; Dufour, Esteban \& Casta\~neda
1996).  The presence of WR stars in the brighter (NW) component
(Izotov et al. 1997a; Legrand et al. 1997; de Mello et al. 1998; Brown
et al. 2002) indicate an ongoing star formation, or a star formation
which has stopped only a few Myr ago.

The age of IZw18 also has been derived from a comparison of the
morphology and the extension of the supershell with simple dynamical
models of superbubble evolution (Martin 1996).  The resulting age
ranges between 15 and 27 Myr, depending on whether the star formation
is continuous or occurs in an instantaneous burst.  The chemical
evolution models of Kunth, Matteucci \& Marconi (1995) were able to
reproduce the C, N, O and Fe abundances of IZw18 by assuming either a
single or two very short bursts (with a duration of no longer than
10-20 Myr), separated by a quiescent period of 1 Gyr.  The integrated
Spectral Energy Distribution of IZw18 can be reproduced either with an
instantaneous burst with an age of $\sim$ 3 Myr, or with a continuous
burst lasting for 13 Myr (Mas-Hesse \& Kunth 1999).  Older stars are
not required.  Takeuchi et al. (2003) tried to reproduce the Infrared
Spectral Energy Distribution observed in IZw18 by means of a dust
production model, based on Hirashita, Hunt \& Ferrara (2002)
prescriptions.  These authors predict an age range between 10 and 40
Myr and do not invoke the presence of older stars.

However, the relatively high C/O and N/O abundance ratios seem to be
justified only by assuming the presence of stars as old as a few
hundred Myr (Dufour, Garnett \& Shields 1988; Garnett et al. 1997)
(note that the C abundance in IZw18 has been recalculated by Izotov \&
Thuan 1999 and seems to be in agreement with the C/O observed in
similar metal-poor BCDs).  Also the remarkable global homogeneity of
the chemical abundances in IZw18 led V\'ilchez \& Iglesias-P\'aramo
(1998) to invoke the presence of an older stellar population, since
otherwise the distribution of metals would be more patchy.  A
breakthrough in the age and SFH determination of IZw18 has come with
the work of ATG.  By means of a careful analysis of Hubble Space
Telescope WFPC2 archival data, these authors were able to detect
post-main sequence stars, identified as Asymptotic Giant Branch (AGB)
stars, with an age of several 10$^8$ yr, as derived from the
comparison of the observed CMD with synthetic ones.  This result has
been confirmed by \"Ostlin (2000), who analyzed the CMD in the near
infrared, revealing the presence of stars as old as a few Gyr.  The
relative importance of this older stellar population compared to the
younger one is however still matter of debate.  The number of red
point sources detected both in the optical and in the NIR ranges by
HST (ATG; \"Ostlin 2000) is not more than a dozen and the Spectral
Energy Distribution of IZw18 is reproduced without the need for stars
older than 13 Myr (Mas-Hesse \& Kunth 1999).  Kunth \& \"Ostlin (2000)
proposed that the low surface brightness component of IZw18 can be due
to a disc of stars as old as 1 -- 5 Gyr, but this finding has been
debated by Papaderos et al. (2002).  These authors found that the low
surface brightness envelope of the galaxy is not due to an old disc
component, but rather to an extended ionized gas emission.

An extreme model has been proposed by Legrand (2000) and Legrand et
al. (2000).  It has been proposed that the chemical and photometric
properties of IZw18 can be reproduced assuming a very low star
formation rate (10$^{-4}$ M$_\odot$ yr$^{-1}$), lasting for a Hubble
time, with a recent burst of star formation superimposed.  More
recently, Hunt, Thuan \& Izotov (2003), by fitting colors of different
regions of IZw18 with evolutionary synthesis models, derived an age of
at most 500 Myr for the oldest stellar population and at most 15 Myr
for the youngest one in the main body.  Stars older than 500 Myr are
not detected.  If any, they would contribute by no more than 22 \% in
mass (4 \% in $J$ light).

In conclusion, the idea that IZw18 is experiencing the very first
burst of star formation is definitively ruled out.  On the other hand,
the recent literature seems to agree on the fact that very old stars
(stars older than 0.5 -- 1 Gyr), if any, do not produce a significant
contribution to the light and the metal budget of IZw18.  We therefore
adopt hereafter the SFH inferred by ATG.

\section{The model}

We simulate a galaxy model resembling IZw18 by means of a 2-D
hydrocode in cylindrical coordinates, coupled with detailed chemical
yields originating from SNeII, SNeIa and winds from Intermediate-Mass
Stars (IMS).  The SNeIa rate is calculated according to the so-called
Single-Degenerate scenario (namely C-O white dwarfs in binary systems
that explode after reaching the Chandrasekhar mass because of mass
transfer from a red giant companion).  This rate can be calculated
analytically for an instantaneous burst of star formation (Greggio \&
Renzini 1983; Matteucci \& Recchi 2001) and it peaks after a few tens
of Myr.  For massive stars we adopt the nucleosynthetic yields of
Woosley \& Weaver (1995), whereas for the IMS we consider both the
yields of Renzini \& Voli (1981: hereafter RV81) and the ones of van
den Hoek \& Groenewegen (1997; hereafter VG97).  For RV81 we chose the
set of yields with Z=0.004 and $\alpha=1.5$, whereas among the
calculations of VG97 we chose the yields with Z=0.001 and
$\eta_{AGB}=4$. In Sect. 4.4 we analyze two models in which new yields
of Meynet \& Maeder (2002; hereafter MM02), for both massive stars and
IMS, are implemented.  The details about the code and the computation
of the chemical enrichment can be found in RMD and R02.

According to the results of ATG, we adopt a star formation constant
for 270 Myr (with SFR = 6$\times$ 10$^{-3}$ M$_\odot$ y$^{-1}$), a
quiescent period of 10 Myr and a second, more vigorous burst (SFR =
3$\times$ 10$^{-2}$ M$_\odot$ y$^{-1}$), lasting for 5 Myr.  Models
with a weaker SFR in the first, long-lasting episode of SF, are
analyzed in Sect. 4.3.

We calculate the mass of stars formed between $t$ and $t$ + $\Delta
t$, where $\Delta t$ $=$ 10$^5$ yr.  This {\it Stellar Population}
(SP) is treated as a starburst.  We thus, according to the adopted IMF
slope and chemical yields, calculate the mass, energy and metal
released by each SP.  The metals released by each SP will vary
according to the metallicity of the stars.  The rate of restitution of
each chemical element $l$ into the ISM from SNeII is calculated as
follows:

\begin{eqnarray}
\dot m^l_{II} (t) & = & \sum_i \phi[m(t-i\cdot \Delta t)]
\biggl[\Delta m(t-i\cdot \Delta t) X^l(i)  + \cr
 & + & m(t-i\cdot \Delta
t)P^l(m)\biggr] \bigg\vert {dM \over dt} 
\bigg\vert_{t-i\cdot \Delta t} 
\end{eqnarray}
\noindent
where the sum is extended to all the SP young enough to contain live stars
more massive than 8 M$_{\odot}$.  We consider in this equation both
the contribution of newly synthesized metals ($m(t-i\cdot \Delta
t)P^l(m)$), and of metals released by the stars but unmodified by
nucleosynthetic processes ($\Delta m(t-i\cdot \Delta t) X^l(i)$).  In
a similar way we calculate the mass loss rate from SNeIa and
intermediate-mass stars.  Analogously to what we have done in RMD and
R02, we consider here the evolution, in space and time, of 8 chemical
elements of particular astrophysical relevance, namely H, He, C, N, O,
Mg, Si, Fe.

We assume a SF occurring at the center of the system and consequently
we inject energy and metals uniformly in this central region.  This
picture is simplistic, since star formation occurs mostly in star
clusters (Meurer et al. 1995; Fall 2004), which can be particularly
large (super star clusters) in dwarf irregular galaxies (Larsen 2002).
However, due to the relatively low resolution of our simulations and
to the assumption of the symmetry along the polar axis, is it not easy
to simulate the sites of star formation and release of energy and
metals in a more accurate way.  We have to take into consideration
this limitation, when comparing the results of our simulation with
IZw18 (see also Sect. 5.1.1).

Our code is able to calculate at each time-step the metallicity of the
star forming region, thus varying the $X^l(i)$ term in Eq. (1) in a
self-consistent way.  However, our simulation cannot spatially resolve
molecular clouds (with sub-parsec dimensions, see e.g. Matsumoto \&
Hanawa 2003).  The calculation of the metallicity of each stellar
population would be obtained just by averaging the metal content of
the cold gas in the star forming region, which is a rather poor
approximation.  Moreover, the superbubble created by the ongoing star
formation soon fills the whole star forming region.  The gas present
in this region is therefore diluted, hot (thus unable to form stars)
and enriched in metals.  For this reason, it is hard to reconcile the
hypothesis of a varying metallicity of the stellar populations with
the centrally concentrated star formation.  We therefore assume, in
this study, a constant metallicity of 1/100 Z$_{\odot}$, according to
the findings of ATG.  This is a simplistic, but not unrealistic,
assumption.

Following the assumptions of RMD, we adopt a low thermalization
efficiency for SNeII ($\eta_{II} = 0.03$; see details in RMD), namely
only 3 \% of the explosion energy of SNe is available to thermalize
the ISM, whereas the rest is radiated away during the expansion of the
Supernova Remnant.  Actually, the first SNe heat and dilute the gas,
thus in principle the thermalization efficiency should increase with
time.  However, SNeII arise from short-living stars (lifetimes $\sim$
3 -- 30 Myr).  These stars are born in molecular clouds, thus always
in a dense and cold environment, irrespective of the energy released
by the other SNe before.  Thus, if SNeII explode, they have to explode
in a medium relatively cold and dense, due to the short lifetime of
massive stars.  If the medium is hot and diluted, stars cannot form
and, of course, SNeII cannot explode.  On the other hand, SNeIa arise
from binary systems with lifetimes spanning from some tens of Myr to
several Gyr.  The progenitor stars are thus born in a cold
environment, but by the time at which the explosion occurs, the
thermodynamical conditions of the circumstellar medium can be
drastically different.  We can therefore safely assume a very high
thermalization efficiency for SNeIa (namely $\eta_{Ia} = 1$).  The
effect of a lower SNeIa thermalization efficiency is analyzed in
Sect. 4.3.  Note that the low thermalization efficiency for SNeII,
disputed in some papers (Strickland \& Stevens 2000; Summers et
al. 2003) has been confirmed, at least in the first $\sim$ 16 Myr of
burst activity, by Melioli \& de Gouveia dal Pino (2003).

Both SNeIa and SNeII are assumed to have an explosion energy of
10$^{51}$ erg.  Some recent papers show a possible dependence of the
explosion energy on the progenitor mass, from both a theoretical
(Fryer 1999; Fryer \& Kalogera 2001) and an observational (Nomoto et
al. 2003) point of view.  However these results contradict each other,
thus no firm conclusion can be drawn about this point and the
assumption of a constant explosion energy seems to be the safest one.

\section{Results}

\subsection{The standard model}

\subsubsection{Dynamical results}

\begin{figure*}[t]
 \begin{center}
 \vspace{-3.5cm}
 \psfig{file=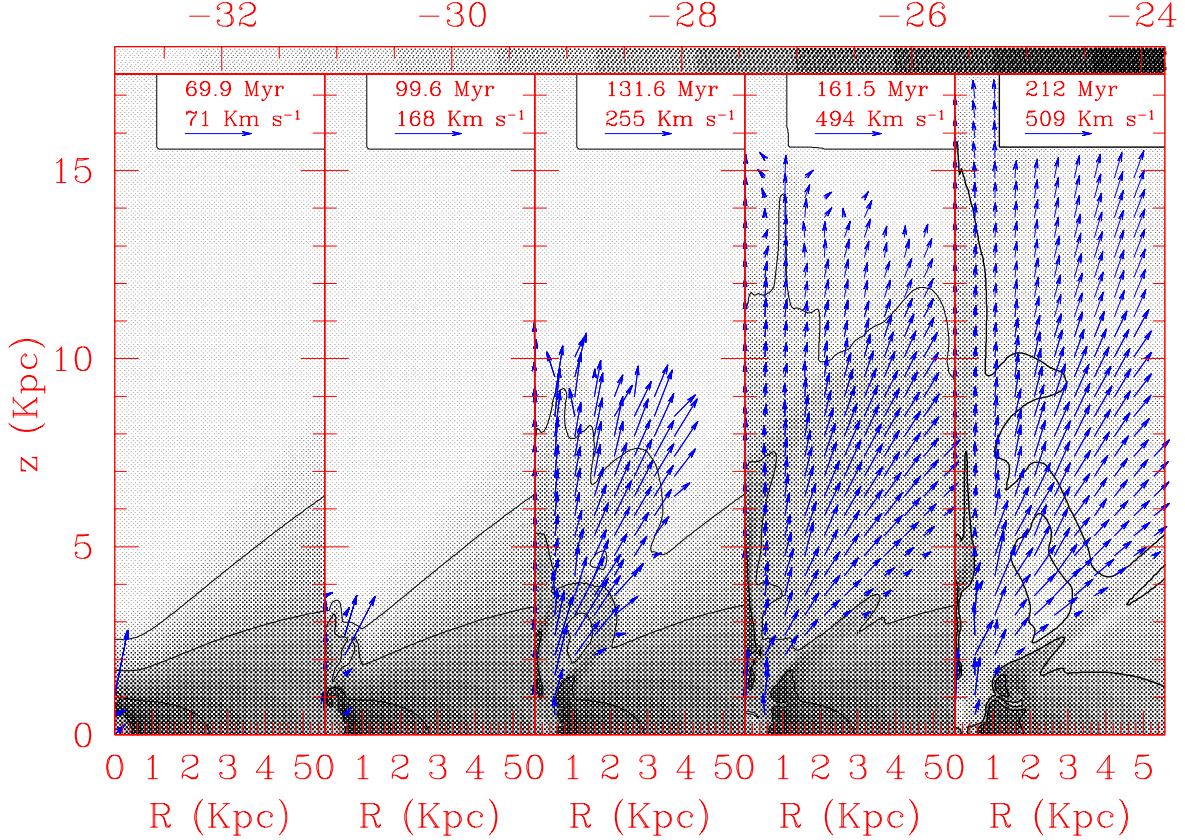,width=15.9cm}
\caption{Density contours and velocity fields for model SR2 at 
different epochs (evolutionary times are labelled in the box on top of
each panel).  The logarithmic density scale is given in the strip on top of
the figure.  In order to avoid confusion, velocities with values lower
than 1/10 of the maximum value (indicated for each panel in the upper
right box) are not drawn.  This is true also for Fig.~\ref{mag1} and 
Fig.~\ref{snap2}.
}
\label{snap1} 
\end{center}
\end{figure*}

The initial conditions chosen for the standard model (SR2) are almost
identical to the starting model assumed in RMD and R02 papers, namely
a rotating gaseous component in hydrostatic isothermal equilibrium
with the galactic potential (sum of a spherical dark halo and a
flattened stellar component) and the centrifugal force.  The dark
matter halo extends over 10 Kpc, whereas the stellar component is
distributed in an oblate structure with dimensions $\sim$ 1 Kpc
$\times$ 700 pc, similar to the distribution of stars observed in
IZw18 (Dufour \& Hester 1990; Papaderos et al. 2002).  This model
assumes a Salpeter IMF (x = 1.35) between 0.1 and 40 M$_\odot$.

In Fig.~\ref{snap1} is shown a snapshot of the evolution of this model
during the first $\sim$ 200 Myr.  The energy released in the gas by
means of stellar winds and SNe is enough to unbind the gas and already
create a galactic wind after $\sim$ 100 Myr.  The mass of stars formed
after 100 Myr is 6 $\cdot$ 10$^5$ M$_\odot$, similar to the mass of
the burst assumed in the standard model of RMD.  Given the similar
assumptions about the thermalization efficiency, the total energy
released by SNeII is thus the same ($\sim$ 2 $\cdot$ 10$^{53}$ erg).
This energy is released on a time-scale three times longer than in the
case of an instantaneous burst of star formation, thus the luminosity
of the burst is three times lower ($L_w \sim 6.5 \cdot 10^{38}$ erg
s$^{-1}$).  The SNeIa contribution however already becomes
non-negligible after a few tens of Myr and after 100 Myr the
luminosity produced by SNeIa is comparable with the luminosity by
SNeII.  Considering also the contribution of stellar winds from
intermediate-mass stars, a jet-like structure is expected from simple
energetical considerations (see RMD).

\begin{figure}[t]
 \begin{center}
 \hspace{3cm} \psfig{file=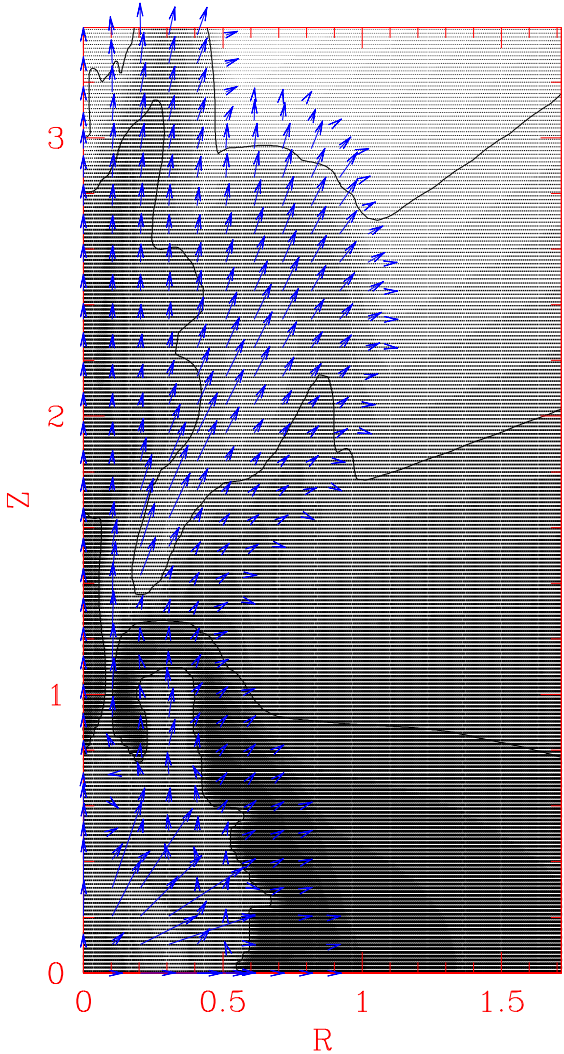,width=15.cm}
\caption{Density contours and velocity fields for model SR2 after 
$\sim$ 100 Myr (magnification of the second panel of Fig.~\ref{snap1}).
}
\label{mag1} 
\end{center}
\end{figure}

A magnification of the second panel of Fig.~\ref{snap1} is shown in
Fig~\ref{mag1}.  In this figure we can observe the classical two-shock
configuration (Weaver et al. 1977): a forward shock is propagating
through the ISM.  Due to the steep density gradient, the shocked ISM
shell accelerates, promoting Rayleygh-Taylor instabilities and
fragmenting into cloudlets and filaments visible in Fig.~\ref{mag1}.
The reverse shock thermalizes the freely expanding wind, heating it up
to temperatures of the order of a few 10$^6$ K.  The reverse shock is
approximately spherical, since the short sound-crossing time (of the
order of 1 Myr) keeps the pressure almost uniform in the inner region.
The extension along the disc is $\sim$ 550 pc.  This is marginally
consistent with the maximal allowed value 0.72 $H_{eff}$
$(L_w/L_b)^{1/6}$ $\sim$ 420 pc, where $H_{eff}$ is the effective
scale-length of the ISM distribution in the polar direction (Koo \&
McKee 1992).  This is also a slightly larger value compared with the
maximal extension in the symmetry plane reached by the model with an
instantaneous burst of star formation (see RMD, their fig. 5),
although the burst luminosity is comparable.  This is apparently at
variance with the findings of Strickland \& Stevens (2000), who show a
slightly slower superbubble evolution when the energy and the mass
injection are more gradual.  A more careful analysis of their fig. 2
shows that the energy injection rate for the ``continuous'' model
(named CSF by the authors) is comparable to the luminosity of the
starburst model only at t $\sim$ 7.5 Myr (the evolutionary time
adopted for the comparison), whereas for t $<$ 7.5 Myr the luminosity
of the CSF model is significantly lower, thus the total energy
injected by SNe in this model is much below the energy released by the
``starburst'' model, at variance with our results.

At the end of the simulation (after $\sim$ 315 Myr), almost half of
the gas initially present inside the galactic region is lost through
the galactic wind.  Only $\sim$ 9 $\cdot$ 10$^6$ M$_\odot$ of gas are
kept bound.  This is a factor of $\sim$ 2 lower than the neutral gas
content of the main body of IZw18, thus this model cannot be the best
one in order to reproduce the morphological properties of IZw18, but
nevertheless it allows us to compare the effect of different star
formation histories (single, instantaneous burst of SF; double burst
of SF; continuous SF) starting from a similar ISM distribution.

\subsubsection{Chemical evolution}

\begin{figure}[t]
 \begin{center}
\vspace{-0.6cm} 
 \hspace{-1cm} \psfig{file=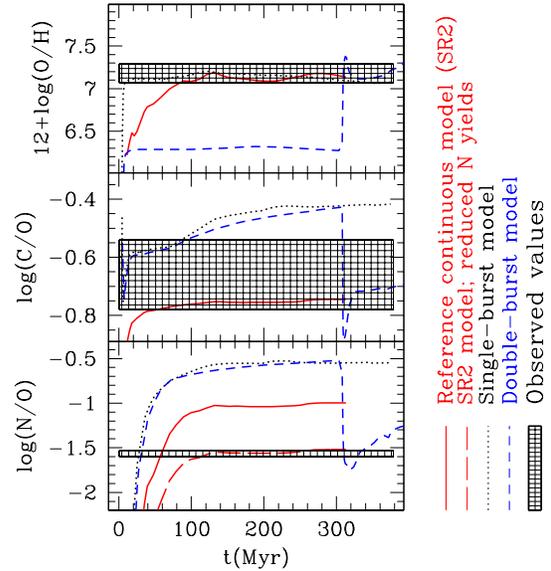,width=9.2cm}
\vspace{-0.3cm}
\caption{ Evolution of O, C, N for the reference continuous burst
model (solid line).  This model is compared with another model with an
instantaneous burst of star formation (dotted line) and a model with
two bursts of star formation separated by a quiescent period of 300
Myr (dashed line).  The long-dashed line is a model in which the N
yields are reduced by a factor of 3.  The superimposed dashed areas
represent the observative values, with relative error-bars, found in
literature for IZw18.  This is true also for Figs.~\ref{cnocont}
and~\ref{cnomm}. }
\label{cnocomp} 
\end{center}
\end{figure}

The evolution of the C, N, O abundances in the cold medium
(i.e. integrating over the grid points in which the temperature is
below 2 $\cdot$ 10$^4$ K) is shown in Fig.~\ref{cnocomp} (solid line).
This evolution is compared with the evolution of similar models (same
initial gaseous distribution, same IMF, same nucleosynthetic
prescriptions, namely the RV81 ones) but with different histories of
star formation.  The dotted line represents the standard single burst
model of star formation (already shown in RMD and labelled as model
M1B therein).  The dashed line represents a model with two
instantaneous bursts of star formation, already shown in R02 (the
model called M300R).

These models show different chemical evolution patterns and it is
worth comparing them to understand the effect of the star formation
history on the chemical evolution of galaxies.  While carbon and
oxygen abundances are reproduced at the end of the simulation of the
continuous burst model, the nitrogen is well above ($\sim$ 0.5 dex)
the values observed in IZw18.  This is due to the fact that the last
burst of star formation is only 5 times more intense than the average
SFR and lasts for only 5 Myr.  The oxygen produced is not enough to
compensate for the nitrogen coming from the IMS.  Moreover, most of
the metals produced in the second episode of star formation do not
have time to cool below 2 $\cdot$ 10$^4$ K, since they are immediately
channelled along the galactic funnel, so they do not contribute to the
chemical enrichment of the galaxy.  At variance with the double-burst
model (dashed line), the signature of the onset of the second episode
of SF is not recognizable.  In order to reproduce the N/O observed in
IZw18, we have to reduce the N yields calculated by RV81 by a factor
of 3 (long-dashed line in Fig.~\ref{cnocomp}).  If we instead adopt
the yields of VG97, a reduction of N yields by a factor of $\sim$ 2 is
enough to fit the observed values of IZw18.

\begin{figure}[t]
 \begin{center}
 \psfig{file=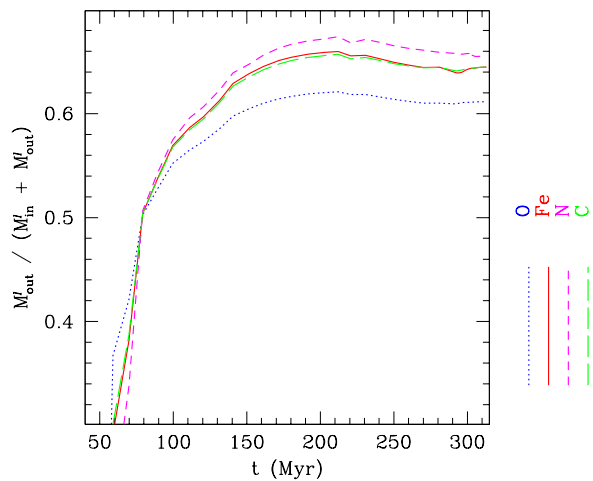,width=8.6cm}
\caption{ Ejection efficiencies ($\eta_l$ = M$^{\it l}_{out}$ / 
(M$^{\it l}_{in}$ + M$^{\it l}_{out}$) ) as a function of time for 
the model SR2.  
}
\label{ejeff} 
\end{center}
\end{figure}

Quite surprisingly, the nitrogen produced by the continuous burst
model during the first, long-lasting episode of star formation, is
well below the value reached by the instantaneous burst models.  This
is only marginally due to the star formation history considered in
this model, and is mostly due to the effect of the galactic wind.  We
can calculate the ejection efficiencies $\eta_l$ by simply dividing
the mass of the gas in the form of the single chemical element $l$
that has left the galaxy (M$^{l}_{\rm out}$) by the total mass of the
chemical species $l$ produced during the episode of star formation,
namely:

\begin{equation}
\eta_l = {{\rm M}^{l}_{\rm out} \over {{\rm M}^{l}_{\rm in} + 
{\rm M}^{l}_{\rm out}}}.
\end{equation}
\noindent
The evolution of the ejection efficiencies of C, N, O and Fe for the
model SR2 are shown in Fig.~\ref{ejeff}.  These four elements are
chosen because they are synthesized by different kinds of stars, on
different time-scales.  Oxygen is mostly produced by quiescent burning
in massive stars and ejected by type SNeII, on a very short
time-scale.  Nitrogen can either be produced in a 'secondary' way,
namely from the C and O already present in the star at birth, or in a
'primary' way, namely stating from the C and O freshly produced in the
star and dredged up during the thermal pulsing phase when the star is
on the AGB phase (RV81).  The secondary production of nitrogen depends
on the metallicity; therefore in metal-poor galaxies the primary
production in intermediate-mass stars is the dominant process.  It has
also been suggested that a significant fraction of N might be produced
in a primary way in massive stars (Matteucci 1986; Izotov \& Thuan
1999), but the N necessary to reproduce the N/O values observed in
metal-poor dwarf galaxies is well above the predictions of
nucleosynthetic papers (see also the discussion in R02).  The carbon
origin has also been debated, since some authors (Carigi 2000; Henry,
Edmunds \& K\"oppen 2000) believe that C is mainly produced in massive
stars.  Chiappini, Romano \& Matteucci (2003a) suggested instead that
C comes mainly from low- and intermediate-mass stars, although the
contribution of massive stars is non-negligible.  Finally, Fe comes
mainly from SNeIa, although SNeII produce around one third of the
observed Fe (Matteucci \& Greggio 1986).

As we can see from Fig.~\ref{ejeff}, the ejection efficiencies become
very large on a time-scale shorter than $\sim$ 100 Myr.  At the
beginning, oxygen is ejected more efficiently than the other elements.
This is due to the fact that the first chemical elements that
experience the effect of the galactic wind are the SNeII products.  At
later times, however, the ejection efficiencies of the elements
produced on longer time-scales become comparable and then slightly
larger than the ejection efficiencies of $\alpha$-elements.  This is
due to the fact that SNeII make the strong effort of breaking-out and
carving the galactic funnel.  When the bulk of SNeIa and
intermediate-mass stars release their nucleosynthetic products into
the ISM, the tunnel has already been carved and it is easy for
chemical elements produced on longer time-scales to be channeled and
ejected out of the galaxy.  This effect has already been found and
analyzed by RMD and R02.  The lower N/O in the continuous model,
observed in Fig.~\ref{cnocomp}, is due in part to the very large
ejection efficiency of N, in part to the fact that SNeII and SNeIa
continuously pump energy into the system, thus the amount of N with a
temperature larger than 2 $\cdot$ 10$^4$ K (thus not considered in
this plot) is larger than in the models with instantaneous bursts of
star formation.

\begin{figure}[t]
 \begin{center}
 \psfig{file=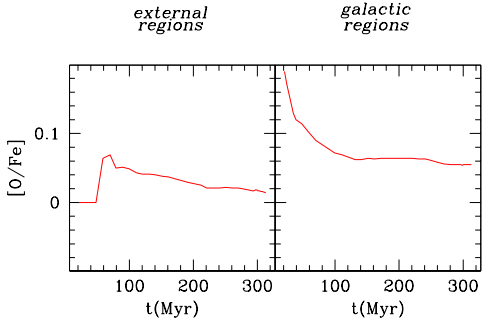,width=6.9cm}
\vspace{-1cm}
\caption{ [O/Fe] vs. time for the model SR2.  
}
\label{afe} 
\end{center}
\end{figure}

The [O/Fe] ratios outside and inside the galaxy are shown in
Fig.~\ref{afe}.  Due to the different ejection efficiencies of O and
Fe seen in Fig.~\ref{ejeff}, there is a small (less than 0.1 dex) but
visible difference in the [$\alpha$/Fe] ratios outside and inside the
galaxy.  This figure also shows a maximum in the [O/Fe] in the gas
flowing outside the galaxy at $\sim$ 70 -- 80 Myr, namely at the
moment of the onset of the galactic wind.  This means that the wind is
initially enriched in $\alpha$-elements, as seen in the outflow of the
starburst galaxy NGC1569 by Martin, Kobulnicky \& Heckman (2002).  The
bulk of Fe in the outflow comes afterwards, due to the minor work
required to eject iron-rich gas when the funnel has already been
carved.

The [O/Fe] shown in Fig~\ref{afe} is still normalized to the Anders \&
Grevesse (1989) solar values, although it is now well established that
these authors overestimated the solar oxygen composition (Holweger
2001).  We keep this value since, due to the difficulties in the
estimation of the iron content in IZw18 (see e.g. Izotov, Thuan \&
Lipovetsky 1997b; Levshakov, Kegel \& Agafanova 2001) and the
uncertainties in the calculation of Fe yields (Timmes, Woosley \&
Weaver 1995) we believe it is meaningless to directly compare the
results of our simulations with the Fe and [$\alpha$/Fe] observed in
IZw18 and thus we concentrate on the comparison between the
[$\alpha$/Fe] inside and outside the galaxy, since it can give us some
hints about the chemical enrichment of the Intracluster Medium.

\subsection{Exploring the parameter space}

As stated in Sect. 4.1.1, model SR2 is not satisfactory to reproduce
the morphological properties of IZw18, since the final gas content is
lower than the observed one.  Moreover, it is possible to reproduce
the abundances of C, N and O at the same time only by reducing the N
yields coming from IMS (see Fig~\ref{cnocomp}).  It is thus worth
changing the initial mass of gas present inside the galaxy, the
adopted set of nucleosynthetic yields and the IMF slope (Salpeter in
the standard model) and explore the parameter space to find the best
model able to fit the chemical and morphological properties of IZw18.
We test here different IMF slopes, different sets of yields from IMS
and different masses of gas at the beginning of the simulation.  In
particular, a flatter (x = 0.5) IMF slope has been suggested by ATG to
reproduce the observed CMD in IZw18.  A steeper (x = 1.7) IMF has been
instead suggested by Scalo (1986) and Kroupa, Tout \& Gilmore (1990)
for massive stars in the Galaxy (see also Kroupa 2002 for a review of
IMF determinations).  The model parameters are shown in Table 1. The
identification of a particular model is made through the notation XYN,
where X refers to the adopted IMF slope (S: Salpeter; K: Kroupa; F:
flat), Y individuates the adopted IMS yields (R: Renzini \& Voli 1981;
V: van den Hoek \& Groenewegen 1997) and N is approximately the gas
mass (in 10$^7$ M$_\odot$) in the galactic region at the beginning of
the simulation.  Therefore, the reference SR2 represents a model with
Salpeter IMF, RV81 yields and approximately 2 $\times$ 10$^7$
M$_\odot$ of \hi gas.

\begin{table}[tp]
\caption{Parameters for the continuous burst models}
\label{model}
\begin{center}
\begin{tabular}{cccc}
  \hline\hline
\noalign{\smallskip}

  Model  &  Gas mass (M$_\odot$) & x (IMF slope) & IMS yields\\
\noalign{\smallskip}

  \hline 
  SR2 & $1.7\times10^{7}$ & 1.35 & RV81    \\
  SR3 & $3.0\times10^{7}$ & 1.35 & RV81    \\
  SV3 & $3.0\times10^{7}$ & 1.35 & VG97    \\
  KR2 & $1.7\times10^{7}$ & 1.7 & RV81    \\
  FV4 & $4.0\times10^{7}$ & 0.5 & VG97    \\
  \hline
 \end{tabular}
\end{center}
\end{table}

\subsubsection{Dynamical evolution}

\begin{figure*}[t]
 \begin{center}
 \vspace{-3.5cm}
 \psfig{file=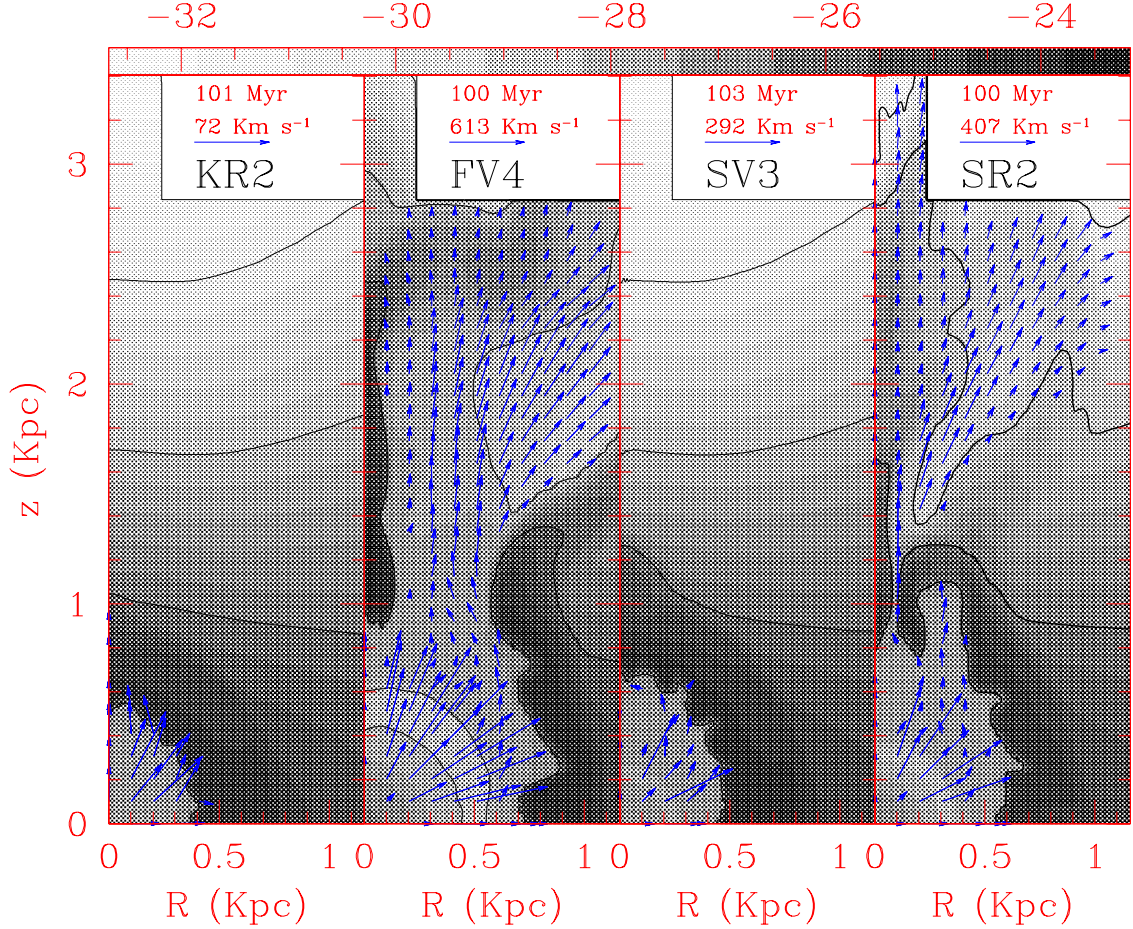,width=15.9cm}
\caption{Density contours and velocity fields for models SR2, SV3,
KR2, FV4 (parameters of these models in Table 1) at $\sim$ 100 Myr.
}
\label{snap2} 
\end{center}
\end{figure*}

As mentioned in Sect. 4.1.1, in the reference model (SR2) a galactic
wind already occurs after $\sim$ 100 Myr (see Fig.~\ref{mag1}).  It is
useful to compare the evolutionary status of the models of Table 1
after $\sim$ 100 Myr, to study how initial gas distribution and the
IMF affect the evolution of the galaxy.  In Fig~\ref{snap2} we show a
comparison of density contours for models SR2, SV3, KR2, FV4.  Model
SR3 is not shown since, from a dynamical point of view, it is almost
indistinguishable from model SV3.

As evident in this figure, only models SR2 (fourth panel) and FV4
(second panel) already develop a galactic wind after 100 Myr.  Model
FV4, in spite of the very high gas content at the beginning of the
simulation, ejects more than 80 \% of the initial mass through the
very powerful galactic wind.  The simulation lasts for $\sim$ 330 Myr
and the final gas content is $\sim$ 7.6 $\cdot$ 10$^6$ M$_\odot$, much
below the observed value.  The flat IMF strongly favors the occurrence
of Type II SNe.  The ratio between the number of SNeII in the FV4
(with a x = 0.5 IMF slope) model and the one in the model with a
Salpeter IMF is simply given by:

\begin{equation}
{N^{FV4}_{SNII} \over N^{SR2}_{SNII}} = 
{\int_8^{40} m^{-1.5} dm \over \int_8^{40} m^{-2.35} dm} \, \simeq 10.
\end{equation}
\noindent
Since we assumed in any models the same prescriptions for the
thermalization efficiencies, the model FV4 produces $\sim$ 10 times
more energy by SNeII than the model SR2.  A difference of a factor of
$\sim$ 4 is also present in the total energy provided by SNeIa.  This
difference is approximately proportional to the ratio between the
number of stars in the range [3, 16] M$_\odot$.  The total energy
provided by SNe in the model FV4 is thus a factor of $\sim$ 7 larger
than the energy provided in the model SR2.  The binding energy of the
gas is instead only a factor of $\sim$ 2 larger and this explains the
remarkable difference in the wind efficiency in the two models.

Model KR2 (first panel in Fig.~\ref{snap2}), with a steeper (x = 1.7)
IMF slope, produces much less energy through SNe (more than a factor
of 2 lower), thus the development of the galactic wind is much
slower.  Only $\sim$ 10 \% of the gas initially present in the galaxy
is lost through the galactic wind, thus the gas content of this model
is consistent with the observations.  Finally, model SV3 (third panel
in Fig.~\ref{snap2}) develops a weak galactic wind at around 170 Myr.
The final gaseous mass is $\sim$ 2.2 $\cdot$ 10$^7$ M$_\odot$, thus
marginally consistent with the gas content observed in IZw18.

\subsubsection{Chemical evolution}

\begin{figure}[t]
 \begin{center}
 \hspace{0.6cm} \psfig{file=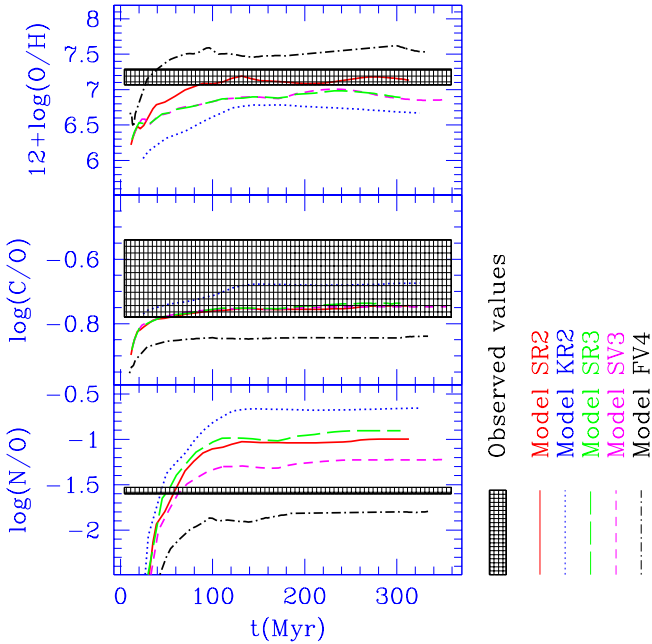,width=9.2cm}
\vspace{-0.3cm}
\caption{ Evolution of O, C, N for the continuous burst models. The 
parameters adopted for each model are described in Table 1.  }
\label{cnocont} 
\end{center}
\end{figure}

The resulting evolution of O, C/O, N/O for the models described in
Table 1 is shown in Fig.~\ref{cnocont}.  As can be seen from the
figure, none of these models can reproduce the observed N/O ratio in a
satisfactory way.  Only the model FV4 (dot-dashed lines in
Fig.~\ref{cnocont}), with a flatter (x=0.5) IMF slope can account for
the low N content (indeed even lower than the observed value), but
this model overproduces O (and, consequently, the predicted C/O is
below the observed values).  Moreover, as described in Sect. 4.2.1,
this model is unable to reproduce the mass content of IZw18.

Model SR2 (solid lines in Fig.~\ref{cnocont}) is the reference model
and it has been already described in Sect. 4.1.2 and shown in
Fig.~\ref{cnocomp}.  Compared to model SR2, model SR3 (long-dashed
lines in Fig.~\ref{cnocont}) contains a dilution of metals produced by
the ongoing star formation in a larger amount of gas, thus resulting
in a lower oxygen abundance.  This is also true for the model SV3
(dashed lines in Fig.~\ref{cnocont}).  The main difference between
model SR3 and SV3 is on the N production.  In fact, O is mostly
produced in massive stars, and a change in the adopted IMS yields does
not influence the O production.  On the other hand, C is produced both
by massive stars and by IMS, with a significant contribution from
masses between 1 and 3 M$_\odot$ (see Chiappini et al. 2003a).  There
is some difference in the C production between RV81 and VG97
nucleosynthetic yields, but the difference arises more clearly after
the lifetime of a 3 M$_\odot$ star, larger than the considered
evolutionary time in our simulations.

Model KR2 (dotted lines in Fig.~\ref{cnocont}), with a steeper (x =
1.7) IMF slope, is strongly biased towards low mass stars, therefore
it overproduces N and underproduces O.  Due to the fact that it
underproduces the C coming from massive stars, as well, it provides a good
fit of the C/O abundance ratio.  Such a very steep IMF in BCD galaxies
can therefore be ruled out.  The best IMF slope in order to reproduce
the characteristics of IZw18 seems therefore to be the Salpeter (1955)
one or flatter (but steeper than x = 0.5).

\subsection{Models with different Star Formation Rates and different
SNeIa thermalization efficiencies}

In order to explore the effect of the variations in the SFH, we ran
two models (labelled SV3b and SV3c), similar to SV3 model, but with
reduced SFR.  The SFR during the first, long-lasting episode of SF is
4 $\times$ 10$^{-3}$ M$_\odot$ y$^{-1}$ (2/3 of the original value)
for the model SV3b and 2 $\times$ 10$^{-3}$ M$_\odot$ y$^{-1}$ (1/3 of
the original value) for the model SV3c.  The SFR of the second, more
recent burst of SF is unmodified.

The injected energy in these two models is strongly reduced, therefore
the fraction of gas able to escape the galaxy is small.  At the end of
the simulation, the amount of gas still present in the galactic region
is $\sim$ 2.8 $\cdot$ 10$^7$ M$_\odot$ for model SV3b and $\sim$ 2.9
$\cdot$ 10$^7$ M$_\odot$ for model SV3c.

\begin{figure}[t]
 \begin{center}
 \hspace{0.6cm} \psfig{file=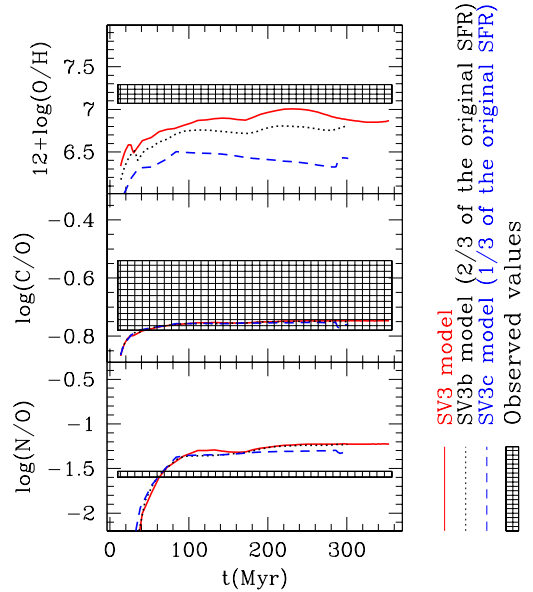,width=9.2cm}
\vspace{-0.3cm}
\caption{ Evolution of O, C, N for the model SV3 (solid line) and for
models with reduced SFR.  Dotted line: model SV3b (SFR during the
long-lasting burst 4 $\cdot$ 10$^{-3}$ M$_\odot$ y$^{-1}$).  Dashed
line: model SV3c (SFR during the long-lasting burst 2 $\cdot$
10$^{-3}$ M$_\odot$ y$^{-1}$).}
\label{cnodiffsfr} 
\end{center}
\end{figure}

The evolution of C, N, O for these 3 models is shown in
Fig.\ref{cnodiffsfr}.  As expected, variations in the SFR do not
appreciably change the C/O and N/O ratios, whereas the global
production of oxygen is reduced.  It is also worth noticing that,
after the onset of the second episode of SF, a bump in the evolution
of O/H in the model SV3c arises.  Such a bump is not observed in the
model SV3 and is only barely visible in the model SV3b.  As pointed
out also in Sect. 4.1.2, in the models with large SFR, most of the
metals produced in the second episode of SF are immediately channelled
along the galactic wind and do not have the chance to cool below 2
$\cdot$ 10$^4$ K, therefore the onset of the second episode of SF is
not observed in the chemical evolution tracks.  Moreover, the mass in
stars formed in the second episode of SF is $\sim$ 8 \% of the total
stellar mass of the galaxy for models with standard SFR.  This
fraction increases up to $\sim$ 12 \% for model SV3b and $\sim$ 22 \%
for model SV3c.  The metal production during the second episode of SF
for these models therefore has a stronger impact on the global metal
budget of the galaxy.

We also ran a model, similar to SV3, but with reduced thermalization
efficiency for SNeIa ($\eta_{Ia} = 0.1$ instead of 1).  In this model
the amount of gas kept bound in the galactic region at the end of the
simulation is $\sim$ 2.8 $\cdot$ 10$^7$ M$_\odot$, significantly more
than the final gaseous mass in the model SV3 ($\sim$ 2.2 $\cdot$
10$^7$ M$_\odot$).  The C, N, O abundances at the end of the
simulations for this model are slightly larger than in the SV3 model,
the difference being of the order of $\sim$ 0.1 dex.  This difference
arises from the fact that more metals are kept bound inside the
galactic region.  Moreover, since less energy per unit mass is
provided by the starburst, the fraction of cold metals is larger.

\subsection{Models with Meynet \& Maeder (2002) yields}

There are indications (Chiappini et al. 2003a) that the sets of yields
of VG97 may overestimate the amount of nitrogen produced by
intermediate mass stars. Recent models of stellar evolution with
rotation (MM02) predict less N in IMS.  MM02 yields were obtained from
self-consistent complete stellar models.  N is mainly produced in a
primary way through rotational diffusion of C in the H burning shell.
We emphasize here that these models do not take into consideration
later phases of the stellar evolution (in particular the third
dredge-up and the hot-bottom burning phase), and may underestimate the
amount of primary nitrogen.  However, the yields of VG97 and RV81 are
obtained by means of synthetic AGB models, in which the production of
N and C, in particular during the hot-bottom burning phase, is
parametrized following some ``ad hoc'' analytical relations, since
none of the available stellar models adequately describe this phase of
the stellar evolution.  Hence, none of the available models in the
literature seems to contain all the physics needed to reproduce the
primary nitrogen production.  Is it therefore interesting to test
different sets of nucleosynthetic yields in order to put constraints
on the primary production of N in IMS.

\begin{figure}[t]
 \begin{center}
\vspace{-0.3cm}
  \hspace{-0.5cm} \psfig{file=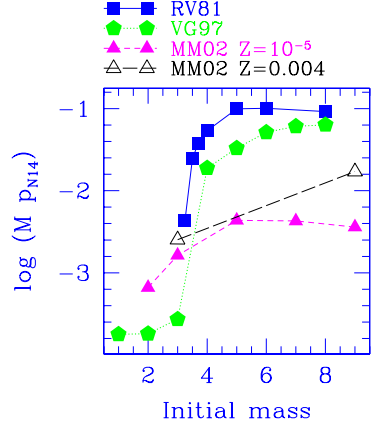,width=5.8cm}
\vspace{-0.2cm} 
\caption{ Nitrogen yields in IMS coming from different authors.
Squares are the results of RV81; pentagons refer to the VG97 yields.
Open triangles show the MM02 models with Z=10$^{-5}$, whereas filled
triangles refer to the models with Z=0.004.  
}
\label{nyields} 
\end{center}
\end{figure}

A comparison of the N yields in IMS coming from different authors is
shown in Fig.~\ref{nyields}.  MM02 presented two sets of yields: the
first one with a very low (Z = 10$^{-5}$) metallicity and the second
one with a metallicity of Z=0.004.  In the case of models with Z =
10$^{-5}$, MM02 do obtain the third dredge-up, in the sense that the
products of H and He burning reach the surface of the star.  As can be
seen in this figure, the predicted production of N in IMS is much
below the values tabulated by RV81 and VG97.  Only at very low initial
masses VG97 yields are below the MM02 results, but this mass range
([1, 3] M$_\odot$) produces a negligible amount of N.  MM02 predict
instead a (slightly) larger amount of primary N coming from massive
stars, compared to the results of Woosley \& Weaver (1995) (see
Chiappini, Matteucci \& Meynet 2003b for details).

We ran two models adopting MM02 nucleosynthetic yields.  For the model
with Z=0.004 only two initial stellar masses (3 M$_\odot$ and 9
M$_\odot$) have been calculated by the authors.  The initial
conditions are the same as the SR3 and SV3 models, i.e. the same
initial gaseous distribution and the same IMF are adopted.  The only
difference is that the IMF extends up to 60 M$_\odot$, since MM02, at
variance with Woosley \& Weaver (1995), consider the yields coming
from stars of initial mass up to 60 M$_\odot$.  Moreover, adopting the
MM02 yields, it is not necessary to mix sets of yields coming from
different authors, since these authors calculated models in the mass
range [2, 60] M$_\odot$.  The results of MM02 have been tested in
chemical evolution models (Chiappini et al. 2003b) and are able to
reproduce the main feature of the Galaxy in a satisfactory way.  We
refer as to MM1 for the model with Z=0.004 and to MM2 for the one with
Z=10$^{-5}$.

\begin{figure}[t]
 \begin{center}
\vspace{-0.3cm}
  \hspace{-0.5cm} \psfig{file=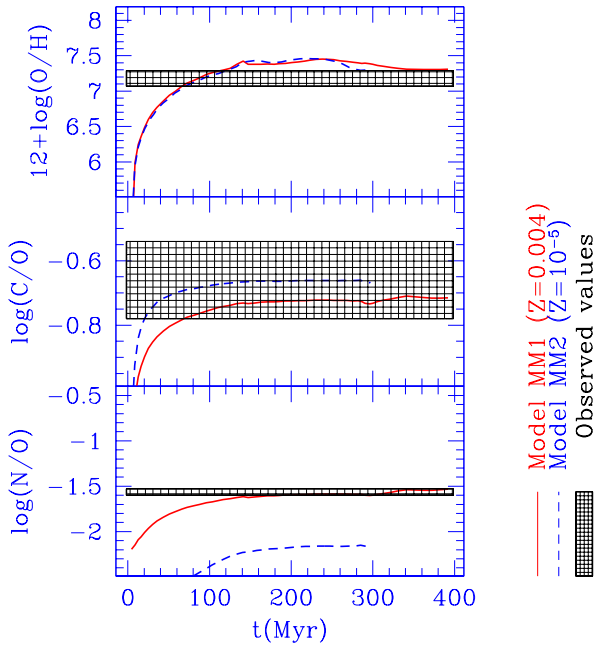,width=9.2cm} 
\vspace{-0.5cm}
\caption{ Evolution of O, C, N for models MM1 (solid line) and MM2 
(dashed line). }
\label{cnomm} 
\end{center}
\end{figure}

The dynamical evolution of models MM1 and MM2 is pretty similar to the
one of model SV3: at the end of the simulation the mass content inside
the galaxy is $\sim$ 2 $\cdot$ 10$^7$ M$_\odot$, thus in agreement
with the observations in IZw18.  The chemical evolution of these two
models is shown in Fig.~\ref{cnomm}.  In spite of the larger
production of N in massive stars, IMS are always required in order to
reach the N values observed in IZw18.  It is thus not possible to
reproduce the observations assuming that only massive stars contribute
to the nitrogen enrichment, as suggested by Izotov \& Thuan (1999; see
also next section).  The N produced in model MM1 at the end of the
simulation, is remarkably in agreement with the observations in IZw18
and, for a considerable range of time, the model is able to reproduce
at the same time the abundances of C, N, O, although the O produced by
this model lies slightly above the observations.  It is however worth
reminding that the MM02 yields do not consider the third dredge-up,
therefore the final N abundance should be considered as a lower limit.
The MM2 model produces instead much less N and it cannot reach the
observed N/O level.  This model can be used to explain the low N/O
observed in some Damped Lyman Alpha (DLA) Systems (see Chiappini et
al. 2003b).

\subsection{The abundances in the \hi medium}

Two recent papers (Aloisi \etal 2003; Lecavelier des Etangs \etal
2004) tried to determine the composition of the \hi gas in IZw18.
These two papers agree in pointing out a (slightly) lower metal
content in the \hi regions compared to the \hii ones, but the final
results of the two papers differ significantly.  A direct comparison
of these two sets of results is shown in Table 2.

\begin{table}[tp]
\caption{Abundances in the \hi medium of IZw18}
\label{model}
\begin{center}
\begin{tabular}{ccc}
  \hline\hline	
\noalign{\smallskip}

  Element ratio &  A03$^a$ & L03$^b$\\
\noalign{\smallskip}

  \hline 

log (\oi/\hi)  & $-5.37$ $\pm$ 0.28 & $-4.7_{-0.6}^{+0.8}$\\
log (\ni/\oi)  & $-1.54$ $\pm$ 0.30 & $-2.4_{-0.8}^{+0.6}$\\
log (\ari/\oi) & $-2.38$ $\pm$ 0.31 & $-3.15_{-0.85}^{+0.6}$\\
  \hline
 \end{tabular}
\end{center}
$^a$ Aloisi et al. (2003)

$^b$ Lecavelier des Etangs et al. (2004)

\end{table}

Aloisi et al. (2003) found an [O/Fe] that was low, although with large
error-bars ([O/Fe] = $-0.3$ $\pm$ 0.3).  This very large iron content
in the neutral medium can be explained only by assuming a significant
iron production in SNeIa, since it is not possible for SNeII to
produce gas with an iron overabundance relative to oxygen.  This means
that SNeIa already contaminated the surrounding gas at the early
phases of the evolution of the galaxy.  Aloisi et al. (2003)
calculated that $\sim$ 470 SNeIa are required in order to justify the
iron content of the \hi medium in IZw18.  This number is well in
agreement with the number of SNeIa exploding in our simulations.  Some
models of Type Ia SN explosion (Kobayashi et al. 1998; Kobayashi;
Tsujimoto \& Nomoto 2000) predict a metallicity threshold for the SNIa
occurrence.  When the metallicity of the gas [Fe/H] is below $-1.1$,
Type Ia SN explosion cannot take place due to strong metal winds in
the primary star.  Actually, the metallicity of IZw18 is well below
this threshold, thus, according to this scenario, SNeIa should not be
possible.

The comparison between these two sets of results show remarkable
differences.  In particular, the O abundance in Lecavelier des Etangs
et al. (2004) is consistent with the O/H ratio observed in \hii
regions, although the error-bars are huge.  Aloisi et al. (2003) show
instead an O abundance a factor of $\sim$ 3 -- 4 lower than in the
ionized gas.  The difference is due to a different choice of \oi
lines: Aloisi et al. (2003) measured the intensity of the line at
$\lambda_0=$ 1039.23~\AA~\ .  Lecavelier des Etangs et al. (2004),
claiming that this line is contaminated by a terrestrial airglow,
chose lines at shorter wavelengths ($\lambda_0=$ 924.95~\AA~\ ,
925.44~\AA~\ , 976.45~\AA~\ ). Nitrogen is depleted by a similar
factor, thus the N/O ratio is consistent with the observations of the
\hii regions.  In Lecavelier des Etangs et al. (2004) instead, due to
the larger O content, the N/O ratio is much below the observations in
the \hii regions.

\begin{figure}[t]
 \begin{center}
\vspace{-0.3cm}
  \hspace{-0.5cm} \psfig{file=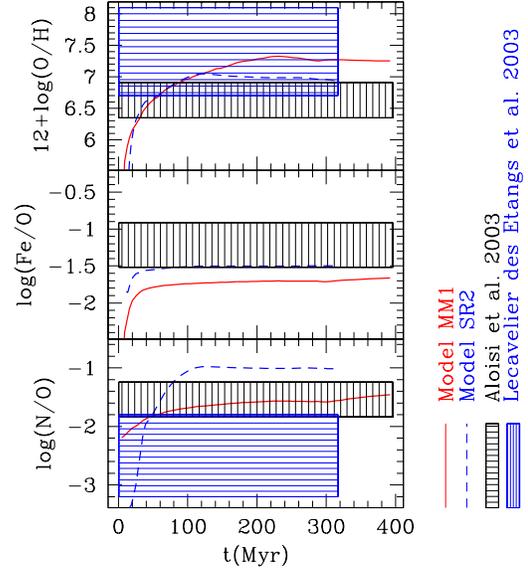,width=9.2cm} 
\vspace{-0.5cm}
\caption{ Evolution of O, Fe, N in the neutral medium for models MM1
(red line) and SR2 (blue line). Vertically dashed area represent
the observations (with relative error-bars) performed by Aloisi et
al. (2003).  Horizontally dashed area refers instead to the
observations of Lecavelier des Etangs et al. (2004).  }
\label{hiabb} 
\end{center}
\end{figure}

We are not able to distinguish clearly the \hi from the \hii regions
in our simulations.  Moreover, since the gas in our model is in
hydrostatic equilibrium with a potential extending out to 10 Kpc, the
abundance ratios in the neutral medium will depend on what fraction of
this gas we decide to consider to calculate the abundance ratios.  We
decided to select all the gas with a temperature below 7 $\cdot$
10$^3$ K in a volume of $\sim$ 2 Kpc $\times$ 2 Kpc in the R-z plane.
The resulting evolution of O, Fe/O and N/O for model MM1 (solid line)
and SR2 (dashed line) is shown in Fig.~\ref{hiabb}, together with the
observations performed by Aloisi et al. (2003; vertically dashed area)
and Lecavelier des Etangs et al. (2004; horizontally dashed area).
The chemical abundances in the \hi and in the \hii medium as predicted
by our models are similar, thus the mixing between these two phases is
quite efficient (see also the discussion in RMD).  The O predicted by
the models seems to be in agreement with the estimates of Lecavelier
des Etangs et al. (2004), thus overestimating the oxygen measured by
Aloisi et al. (2003).  Only Aloisi et al. (2003) measured the iron
abundance, which is, as pointed out above, overabundant compared to O.
Our models predict an [O/Fe] ratio close to solar.  Finally, the
predictions of model MM1 are in agreement with the estimates of the
N/O ratio in the \hi regions by Aloisi et al. (2003), whereas the
measurements of Lecavelier des Etangs et al. (2004) are well below
these predictions.

\section{Discussion}

\subsection{Physical and numerical limitations}

This paper presents the detailed chemical and dynamical evolution of a
galaxy ongoing a complex star formation history. It moreover includes
a detailed treatment of SNeIa contribution.  Few of previous similar
studies of star-bursting dwarf galaxies take into consideration these
effects.  They usually assume a very simple star formation history (in
most of the cases an instantaneous burst of star formation) and often
neglect the contribution of SNeIa.

Our study is thus a progress in this field, but nevertheless, to
properly discuss the results of our simulations, it is important to
analyze the limitations of the models.  We can distinguish two kinds
of limitations: those related to the inaccuracy in reproducing the
initial conditions in IZw18 and those related to the code itself, in
particular missing physics and numerical resolution problems.

\subsubsection{Initial conditions}

As described in Sect. 3, our model is a 2-D one in which cylindrical
coordinates are assumed.  The initial gaseous distribution can be seen
in RMD, their fig. 1.  This distribution and the way in which the
input of energy and metals is considered cannot exactly reproduce the
real characteristics of IZw18 for several reasons:

\begin{itemize}

\item the distribution of stars in IZw18 is not axially symmetric, as
described in Sect. 2.  Most of the young stars are concentrated in two
star forming regions, whose centers are separated by more than 300 pc
(at a distance of 12.6 Mpc, cfr. \"Ostlin 2000).  Two \hii regions are
associated to this star forming regions.  It is impossible to
accurately reproduce this ``peanut-shaped'' morphology with a 2-D
numerical code.

\item Also the \hi distribution in IZw18 is strongly non-axially
symmetric, with an elongated main component with size 2.2
$\times$ 2.9 Kpc and many other patchy structures surrounding this main
component (van Zee et al. 1998).  Our models cannot reproduce this
very complex \hi distribution.

\item The energy input rate is very simplified in our model.  The
energy input rate calculated by the Starburst99 model (Leitherer et
al. 1999) shows a much more complex behaviour.

\item Wolf-Rayet (WR) stars, although observed in IZw18 (Legrand et
al. 1997; de Mello et al. 1998; Brown et al. 2002) are not considered
in our simulations.  Strong stellar winds coming from these stars may
change the energy and metal input rate.

\item The energy and metal injection occurs uniformly in a central
region.  The effect of discrete explosions is neglected (see
discussion in RMD) and the possibility of inhomogeneous mixing (Argast
et al. 2000) is also excluded.  Actually, the metallicity gradient in
IZw18 is almost flat in the inner 400 -- 500 pc (Legrand 2000), thus
the mixing seems to be rather homogeneous.

\end{itemize}

\subsubsection{Missing physics}

\begin{itemize}

\item Radiative transfer is not included in our code, thus the
distinction between the different ISM phases (\hi and \hii regions) is
made on the basis of the temperature of the medium.

\item A smooth ISM distribution does not represent the correct
description of the ISM in galaxies.  A significant fraction of gas is
in a cloudy phase, which can interact with the diffuse ISM phase
through condensation-evaporation processes, changing the thermal
balance and possibly also the chemical evolution patterns (Theis,
Burkert \& Hensler 1992; Rieschick \& Hensler 2000).  Models including
a cloudy phase will be presented in a forthcoming paper (Recchi et
al. 2004 in preparation).

\item A dusty phase is not included in our code.  In spite of the low
metallicity content of IZw18, the observed amount of dust is of the
order of 2000 -- 5000 M$_\odot$ (Cannon et al. 2002), thus a
significant fraction of metals is stored in the dusty phase.  The
metals observed in the \hii and in the \hi regions can therefore be
only a lower limit of the total amount of metals produced in IZw18.

\item Self-gravity is not included in our code.  It is assumed that
the gravitational potential is dominated by the dark matter halo
(fixed in time).  Self-gravity may change the behaviour of swept-up
shells, due to their large densities.  Gravitationally-induced
fragmentation may occur, possibly producing star formation in the
shell, as suggested by many authors (Larsen, Sommer-Larsen \& Pagel
2001; Ehlerov\'a \& Palous 2002).

\item Contact discontinuities (the interface between the shocked wind
and the shocked ISM) are not properly resolved in any numerical
scheme, due to the well-known numerical diffusion problem.  In our
code the contacts are spread over 4 -- 8 zones.  Inaccuracy in the
treatment of contacts and poorness in the spatial resolution (in our
code the central resolution is 5 pc and the size ratio between
adjacent zones is 1.03) may lead to wrong determinations of the
cooling time-scale of metals.  Actually, recent hydrodynamical
simulations (de Avillez \& MacLow 2002; Marcolini, Brighenti \&
D'Ercole 2004) show that the amount of cooled ejecta is only poorly
sensitive to the numerical resolution of the simulation.  A discussion
of this point is also addressed in Sect. 3.2 of RMD.

\end{itemize}

Some of the limitations will be resolved in forthcoming papers, some
others are harder to treat.  However, it is worth noting that IZw18
shows almost uniform properties (i.e. no significant metallicity
gradients, marginal uniformity in the metallicities of the NW and of
the SE region; see e.g. V\'ilchez \& Iglesias-P\'aramo 1998; Legrand
2000), thus the comparison of the results of our models with the
global properties of the main body of IZw18 is meaningful.

\subsection{The evolutionary history of IZw18}

What can we infer about the evolutionary history of IZw18 from these
chemical and dynamical simulations?  First of all in our models we do
not need stars older than 300 Myr to justify the main chemical and
dynamical properties of IZw18.  Observations of the stellar
populations in nearby Local Group galaxies show the presence of old
stars (i.e. with an age of several Gyr) in all the dwarf galaxies, of
both late and early type (see e.g. Grebel 1997).  In the galaxies in
which the spectroscopy and photometry is deep enough, the presence of
old stars is thus ubiquitous.  If we go further, outside the Local
Group, we start looking at objects in which the presence of older
stars is more questionable.  One of the best examples of young
galaxies is SBS 0335-052, a nearby, very metal-poor BCD galaxy.  Vanzi
et al. (2000) showed that the NIR emission is dominated by very young
(t $\simlt$ 5 Myr) stars and no need for older stars has been invoked.
NGC 1569, another well-known nearby dwarf galaxy, is a post-starburst
galaxy dominated by a recently formed, perhaps still forming, star
cluster system (Anders et al. 2004), although evidence for the
presence of older stars has been observed (Vallenari \& Bomans 1996;
Greggio et al. 1998; Aloisi et al. 2001).

As described in Sect. 2, the age of IZw18 has been questioned for many
years.  It is nowadays accepted that IZw18 is not a galaxy
experiencing star formation for the first time.  An intermediate-age
population of stars is present.  Optical spectroscopy and photometry
is not deep enough to resolve stars on the main sequence, allowing us
only to go down to a look-back time of the order of 1 Gyr (ATG;
\"Ostlin 2000), so stars older than 1 Gyr, if any, would be
undetected.  As described in Sect. 2, the method of color fitting
developed by Hunt et al. (2003) allows the authors to put constraints
on the mass of stars older than $\sim$ 500 Myr.  These stars do not
contribute more than 22 \% to the total mass in stars of IZw18.

From a chemo-dynamical point of view we note that an episode of star
formation lasting for $\sim$ 300 Myr produces a very large amount of
N, synthesized in a primary way in intermediate-mass stars.  As can be
seen in Fig.~\ref{cnocont}, the N/O ratio in a continuous burst model
is almost stable after $\sim$ 100 Myr.  This is due to the fact that
the nitrogen continuously produced by the model is ejected outside the
galaxy more easily than the oxygen, therefore it does not accumulate
inside the galaxy.  Only a very long quiescent period can completely
quench the O production, resulting in an increase of the N/O ratio, up
to the moment in which a second burst of star formation releases a
large amount of O into the ISM, thus drastically reducing the N/O (see
Fig.~\ref{cnocomp}, dashed line).  An episode of star formation
lasting more than 300 Myr is therefore acceptable from the chemical
point of view.

From a dynamical point of view, in spite of the low level of star
formation, the energy produced and restored into the ISM by SN
explosions and stellar winds is very large and able to unbind a
significant fraction of the ISM.  As described in Sect. 3, the energy
feedback depends on some poorly constrained parameters, in particular
on the thermalization efficiency.  Our values of the thermalization
efficiencies are calculated in a self-consistent way starting from the
average values of temperature and density in the region in which the
feedback occurs.  Nevertheless, the physics of the interaction of the
Supernovae with the surrounding ISM is poorly known and the
thermalization efficiency should be considered as a free parameter.
The adopted thermalization efficiency for SNeIa is large and in the
``real world'' this value can be also significantly lower (although
the average explosion energy of SNeIa should be a little larger than
the adopted value of 10$^{51}$ erg).  As seen in Sect. 4.3, a very low
SNeIa thermalization efficiency helps in keeping a larger fraction of
ISM bound in the galactic potential well.  The adopted thermalization
efficiency for SNeII is instead very low.  Strickland \& Stevens
(2000) for example adopted a thermalization efficiency equal to unity
for SNeII.  Thus in a galaxy like IZw18, a constant star formation
rate, even at a low level, cannot be sustained for more than a few
hundred Myr, unless a very low thermalization efficiency for both
SNeII and SNeIa is adopted.  A flatter IMF, as suggested by ATG, would
result in a much higher energy production rate, so it would easily
unbind the gas in the galaxy.  A possible alternative solution is an
upper cut-off in the IMF mass range well below 40 M$_\odot$ (Kroupa \&
Weidner 2003; Tolstoy \& Venn 2003).  A paper based on a set of models
of this kind is in preparation.

\section{Conclusions}

In this paper, we have computed the chemical and dynamical evolution
of a galaxy similar to IZw18 under the assumption of a continuous,
weak burst of star formation lasting for 270 Myr and a 5 times
stronger burst occurring more recently and lasting for 5 Myr.  Such a
star formation history has been suggested by Aloisi et al. (1999) in
order to reproduce the observed CMD of IZw18.  We have tested
different IMF slopes and different sets of yields for
intermediate-mass stars and massive ones.  Different initial gaseous
distributions are also tested.  In most of the explored cases, a
galactic wind develops, as a consequence of the energy produced by
SNeII, SNeIa and winds from intermediate-mass stars.  The time-scale
for the development of the galactic wind depends mostly on the IMF
slope which can change the energy input rate by an order of magnitude.
The mass of gas initially present inside the galaxy influences also
the development of a galactic wind, but not significantly.

Our main conclusions can be summarized as follows:
\begin{itemize}

\item
At variance with models with instantaneous bursts of star formation,
in which the variations in the abundance ratios occurred on very short
time-scales, thus matching the observed values of IZw18 only for small
intervals of time (see dotted and dashed lines in Fig.~\ref{cnocomp}),
in these continuous star formation burst models, the chemical
composition of the gas is stable for longer time-scales.  However,
models with a Salpeter (1955) IMF, when the IMS yields of van den Hoek
\& Groenewegen (1997) or Renzini \& Voli (1981) are implemented,
overestimate the N composition of the ISM by a factor of 2 -- 3.  It
is therefore possible to match the observations of the N/O ratio in
IZw18 by reducing the yields from the above authors by the same factor
(see long-dashed line in Fig.~\ref{cnocomp}).  Due to the
uncertainties in the computation of the N yield, this reduction cannot
be ruled out.  Models with a steep (x = 1.7) IMF largely overestimate
N and underestimate O.  Finally, models with a flat (x = 0.5) IMF
slightly overestimate O and underestimate N.  These flat IMF models,
although hardly acceptable from a dynamical point of view (they inject
too much energy into the gas), can be fine tuned and can be the most
promising models, if Renzini \& Voli (1981) or van den Hoek \&
Groenewegen (1997) yields are implemented, in order to reproduce the
chemical properties of IZw18.

\item
New models of IMS and massive stars evolution with rotation (Meynet \&
Maeder 2002) predict a larger N primary production in massive stars.
Since they do not reach the hot-bottom burning phase, they also have a
significantly reduced production of nitrogen in IMS.  By means of
these yields, it is possible to reproduce the abundances of the ISM in
IZw18 in a satisfactory way.

\item
Our present models confirm the existence of the so-called {\it
differential winds}, namely the freshly produced metals are ejected
more easily than the pristine gas.  In agreement with the results of
Recchi et al. (2001), the metals with the largest ejection
efficiencies are the ones produced on longer time-scales (Fe and N in
particular), since, when the bulk of these metals is produced, a
galactic tunnel has been already carved and it is easy for the
galactic wind to carry them outside the galaxy.  From a chemical point
of view, the consequence of this selective wind is an [$\alpha$/Fe]
ratio in the gas outside the galaxy lower than in the gas inside.
This difference is evident but much smaller than the one produced in
models with a single, instantaneous burst of star formation (Recchi et
al. 2001). Nevertheless, also in this case SNeIa, often ignored in
previous similar works, play a decisive role in the evolution of BCD
galaxies.  This selective wind can also explain the large amount of Fe
and the very low [$\alpha$/Fe] ratio observed in the Intracluster
Medium, if the same mechanism acts in elliptical galaxies which are
the main contributors to metals in the ICM (see Pipino et al. 2002).

\item The chemical composition of the \hi gas can also be inferred
from our models.  The resulting abundances and abundance ratios are
fairly well in agreement with the results obtained in the \hii
regions, showing that the mixing between these two phases is efficient
and fast.  In fact, mostly due to the low thermalization efficiency of
SNeII, the cooling time-scale of freshly produced metals is short (of
the order of a few tens of Myr).  The ``instantaneous mixing''
assumption, widely adopted in chemical evolution models, is therefore
acceptable.  The results of the chemical evolution in the \hi medium
can be compared with recent observations of the \hi regions of IZw18
performed with {\sl FUSE} (Aloisi \etal 2003; Lecavelier des Etangs
\etal 2004).  This comparison does not allow us to impose robust
constraints, since the results of these two groups are remarkably
different.

\end{itemize}

\begin{acknowledgements}

We thank the anonymous referee for many comments and suggestions which
improved the paper.  We are grateful to Cristina Chiappini for useful
comments and for having provided us interpolations to the yields of
Meynet \& Maeder (2002) and Gerhard Hensler for stimulating
discussions.  S.R. acknowledges generous financial support from the
Alexander von Humboldt Foundation and Deutsche Forschungsgemeinschaft
(DFG) under grant HE 1487/28-1.  F.M., S.R and A.D'E. acknowledge
financial support from the INAF (Italian National Institute for
Astrophysics) contract PRIN2002 by the title ``Blue Compact Galaxies:
primordial helium and chemical evolution''.
\end{acknowledgements}

\end{document}